\documentclass[aps,pra,twocolumn,showpacs,floatfix]{revtex4}

\usepackage{bm}
\usepackage{graphicx,psfrag}
\usepackage{amsmath}
\usepackage{amssymb}
\usepackage{color}
\def\beq{\begin{equation}}
\def\eeq{\end{equation}}
\usepackage{latexsym}
\usepackage{amsfonts}

\begin{document}

\title{Two-dimensional expansion of a condensed dense Bose gas}

\author{{E. S. Annibale}$^{1,2}$}
\email{annibale@if.usp.br}
\author{A. Gammal$^{1}$}
\email{gammal@if.usp.br}
\author{K. Ziegler$^{2}$}
\email{klaus.ziegler@physik.uni-augsburg.de}

\affiliation{
$^1$Instituto de F\'isica, Universidade de S\~ao Paulo,  05508-090, S\~ao Paulo, Brazil\\
$^2$Institut f\"{u}r Physik, Universit\"{a}t Augsburg, D-86135, 
Augsburg, Germany}

%\date{\today}

\begin{abstract}

We study the expansion dynamics of a condensate in a strongly interacting Bose gas 
in the presence of an obstacle. Our focus is on the generation of shock waves after 
the Bose gas has passed the obstacle.
The strongly interacting Bose gas is described in the slave-boson representation.
A saddle-point approximation provides a nonlinear equation of motion for the macroscopic wave function,
analogous to the Gross-Pitaevskii equation of a weakly interacting Bose gas but 
with different nonlinearity.  
We compare the results with the Gross-Pitaevskii dynamics of a weakly interacting
Bose gas and find a similar behavior with a slower behavior of the strongly
interacting system. 
\end{abstract}

%------------------

\maketitle
%%%%%%%%%%%%%%%%%%%%%%%%%%%%%%%%%%%%%%%%%%%%%%%%%%%%%%%%%%%%%%%%%%%%%%%%%%%%%%%%%%%%%%%%%%%%%%%%%%%%
\section{Introduction}

Many interesting features of a trapped Bose gas are described by the Gross-Pitaevskii 
equation. This includes dynamical studies of the Bose-Einstein condensate (BEC), such as the expansion 
of a BEC 
after switching off the trapping potential or the interference of two separate BEC's, has been performed 
with remarkable accuracy \cite{RMP1999}.
The Gross-Pitaevskii (GP) equation provides the macroscopic quantum state of the condensed part of 
the Bose gas. It can
be derived from a microscopic statistical model of many-body states at temperature $T$ in the limit 
$T=0$ by a saddle-point approximation.
However, the GP equation is restricted to a dilute Bose gas, with less then one particle
in the scattering volume. More recent experiments with optical lattices have revealed that a much
richer physics appears in a dense (or strongly interacting) Bose gas \cite{RMPlattice}. 
The idea is that a static
periodic potential of the optical lattice is provided by counter-propagating Laser fields, where 
particles occupy the local minima
of the periodic potential. As soon as there is more than one particle per minimum the Bose gas
must be considered as strongly interacting. An immediate effect of stronger interaction is the
depletion of the condensate caused by the collision of particles. This is also known from the
famous example of interacting bosons in form of superfluid helium. If the density of the Bose gas
increases further we can even destroy the condensate completely and create a new quantum state in the
form of a Mott insulator. In contrast to a BEC, the Mott insulator is characterized by local
conservation of the particle number (i.e. $n=1,2,...$ particles per minimum of the optical lattice)
but without phase coherence.
In a trapped Bose gas, contrary to a translational invariant Bose gas, both states can co-exist in
the same system, which is known as the wedding-cake structure: at commensurate densities the system
is in a Mott state with constant density and at incommensurate densities the system is in a Bose-Einstein
condensed state with spatially changing density.

These strongly interacting systems cannot be described within the GP approach because the latter only 
takes 
into account the condensate and neglects the interaction with non-condensed particles. In particular, 
depletion of the condensate or the formation of a Mott insulating state is not accessible by the GP 
approach. Fortunately, there is an extension of the GP equation that is able to take into account the
interaction with the non-condensed particles. It is known as the slave-boson (SB) approach and provides
also a nonlinear Schr\"odinger equation for the macroscopic quantum state of the condensed part of 
the Bose gas. The interaction with the non-condensed particles leads to a modified nonlinearity though.
For a dilute Bose gas the nonlinearity is the same as that of the GP equation but in the regime of the 
dense Bose gas it is weaker than that of the GP equation. The equation of the strongly interacting 
Bose gas also indicates the disappearance
of the condensate when we approach the Mott-insulating state at higher density. Previous studies of
a dense trapped Bose gas have shown that the BEC can be completely destroyed at the trapping center
due to depletion at higher densities \cite{pra56_1438,ziegler02}. On the other hand, the vortex structure
is not much affected in the strongly interacting system because the density of the condensate is low
in the vicinity of the vortex core \cite{pra78_013642}.  

The two-dimensional expansion of a dilute BEC past an obstacle is a subject of intensive theoretical and 
experimental studies. A relevant parameter in these studies is the Mach number $M$. It is 
defined as the ratio of the asymptotic velocity of the flow to the sound velocity in the medium 
\cite{def_sound}. For subsonic Mach numbers in the range $0.5 \lesssim M \lesssim 0.9$, it was reported 
the generation of pairs of vortices and antivortices \cite{frisch}. In case of a supersonic flow, the 
generation of oblique dark solitons inside, and Kelvin ship waves outside the Mach cone (imaginary lines 
drawn from the obstacle at angles $\pm arcsin(1/M)$ with the horizontal axis) were found 
\cite{winiecki_prl,winiecki,prl97_180405,pra75_033619,npa790}. It was shown in \cite{anatoly} that these 
dark solitons are convectively unstable for large enough flow velocity ($M \gtrsim 1.5$), i.e., 
practically stable in the region around the obstacle. 
\linebreak
A general theory on dispersive shock waves for 
supersonic flow past an extended obstacle was developed in \cite{pre} and a review paper on dark solitons in 
BEC can be found at \cite{frantzeskakis}.  Experiments addressing this problem were described in 
\cite{cornel,carusotto,neely10}. Recently, a renewed theoretical interest in this issue was brought by the 
observation of an alternating vortex emission for a suitable set of parameters. These are analogous to the 
``von K\'arm\'an vortex street'' in classical dissipative fluids \cite{saito2010}.

In this paper we shall study the effect of the interaction between the BEC and the non-condensed
particles in a dynamical situation, where a BEC is released from a parabolic trap and passes an 
obstacle. The obstacle is modeled by an impenetrable disc.
Due to a complex interference the macroscopic wave function will experience strong density
fluctuations. The results of our numerical simulation, based on the strongly interacting gas (SIG)
equation, will be compared with previous calculations, based on the GP equation.
The paper is organized as follows: We briefly introduce the SIG equation and compare it with
GP equation. 
%(For its derivation we refer to the existing literature.)
Then we present the results of our numerical simulation for an expanding cloud 
in two dimensions that passes an obstacle. Finally, we discuss these results and compare them with
those of the GP approach.
\begin{figure}[t]
\begin{center}
\includegraphics[width=7cm]{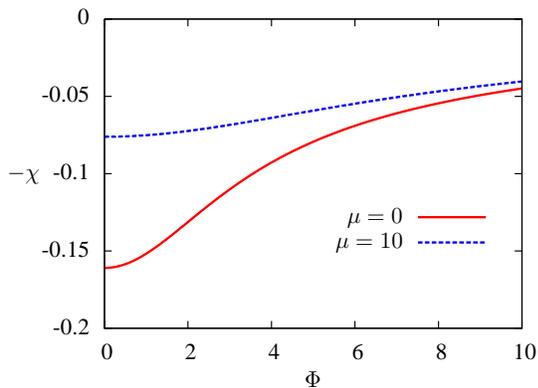}
\caption{\label{chi1}  Behavior of the nonlinear term $-\chi$ in Eq. (\ref{pra8}) 
as a function of $|\Phi|$
for $\mu = 0$  and $\mu = 10$.}
\end{center}
\end{figure}
\begin{figure}[t]
\begin{center}
\includegraphics[width=8cm]{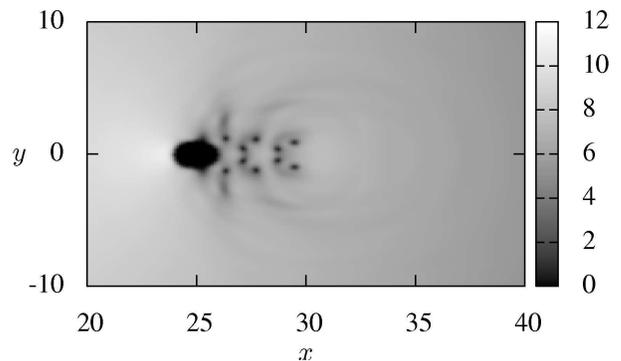}
\caption{\label{vortices}  Numerical solution of Eq. (\ref{sb}) for the initial
profile given in Eq. (\ref{ic}) with $\sigma_1 = 16$ and $\sigma_2 = \sqrt{3} \sigma_1$. 
The density plot is shown for $t = 2$ and $\mu = 0$.
We observe the generation of pairs of vortices and antivortices past the obstacle.
An impenetrable disc of radius $r = 1$ is placed at $(25,0)$ in a BEC radially expanding from
the centre $(0,0)$.}
\end{center}
\end{figure}

%%%%%%%%%%%%%%%%%%%%%%%%%%%%%%%%%%%%%%%%%%%%%%%%%%%%%%%%%%%%%%%%%%%%%%%%%%%%%%%%%%%%%%%%%%%%%%%%%%%%
\section{Slave-Boson Approach}

We start from a Bose gas with hard-core interaction of a given radius $a$. 
Then the Bose gas can be approximated by a lattice gas with lattice constant $a$. 
In other words, $a$ provides the shortest relevant length scale in our Bose gas.
This scale remains in the Bose gas even after its release from the trap. 
In general, a strongly interacting Bose gas has two constituents, namely condensed 
particles and non-condensed particles. This is the case even at zero temperature.  
The interplay of all these particles can be described by the slave-boson approach
\cite{pra56_1438, ziegler02, jphys40_629, annphys17_48, pra78_013642}. Although this
is a many-body picture, the macroscopic wave function of the condensate is extracted
by a variational procedure with respect to the density of the condensate 
analogous to the derivation of the GP equation from the weakly interacting
many-body Bose gas model. The corresponding effective Hamiltonian of the macroscopic 
condensate state $\Phi({\bf r})$ % in a $d$-dimensional Bose gas 
is
\beq
H_\Phi=~
-\frac{J a^2}{6} \, \Delta % \Phi({\bf{r}}) 
\, + \,
J \left (
\alpha_1 - \alpha_2 \chi  \right)
\ ,
\label{hamiltonian0}
\eeq
where $\Delta$ is the three-dimensional Laplacian. 
Moreover, $\chi$ is a term that is nonlinear in $|\Phi|^2$:
\begin{eqnarray}\label{pra8}
\chi({\bf{r}})&&=
\frac{\partial \log Z({\bf{r}})}{\partial | \Phi({\bf{r}}) |^2}=\nonumber\\
&&\frac{1}{2} \frac{1}{Z({\bf{r}})} \int_{-\infty}^{\infty}
e^{- \varphi^2} \left (
\frac{\cosh \gamma({\bf{r}})}{\gamma({\bf{r}})^2} 
- \frac{\sinh \gamma({\bf{r}})}{\gamma({\bf{r}})^3} \right)d \varphi
\ , \nonumber\\
\end{eqnarray}
where $Z({\bf{r}})$ is the integral expression
\begin{equation}\label{pra4}
Z({\bf{r}})~=~\int_{-\infty}^{\infty} 
e^{- %\beta' 
\varphi^2} \,
\frac{\sinh \gamma({\bf{r}})}{
\gamma({\bf{r}})}d \varphi
\end{equation}
with
\begin{equation}\label{pra9}
\gamma({\bf{r}})~=~\sqrt{\bigl( \varphi \, + \, \mu/2 \bigr)^2 \, + \, 
|\Phi({\bf{r}})|^2} \hspace{0.2cm} .
\end{equation}
The coefficients are $\alpha_1=1+\alpha$, $\alpha_2=\alpha(1+1/\alpha^2)$, where
$\alpha$ is a numerical constant $\alpha\approx 1/5.5$ \cite{pra56_1438, pra78_013642}.
The kinetic parameter $J$ is associated with the mass of the bosons $m$ by the relation
\[
\frac{J a^2}{6}=\frac{\hbar^2}{2m}
\ .
\]
The dimensionless parameter $\mu$ is a one-particle chemical potential that is associated
with the density of bosons. This can be understood as a potential that controls the exchange
of particles with the gas outside the trapped cloud by assuming that the cloud is a grand-canonical
ensemble of atoms. Then the value of $\mu$ fixes the number of bosonic atoms in equilibrium.

The obstacle is included by choosing specific boundary conditions for the macroscopic
wave function. In our case this is a hard disc, where the wavefunction vanishes inside
the disc. Finally, the number of condensed bosons $N_0$ is determined by an integral 
of $|\Phi({\bf{r}})|^2$ 
over the entire volume of a three-dimensional Bose gas as
\begin{equation}\label{density}
N_0~=~\frac{1}{a^3} \frac{1}{(1 + 1/\alpha)^2} \int |\Phi({\bf{r}})|^2 d^3 r \hspace{0.2cm} .
\end{equation}
%\begin{figure}[t]
%\begin{center}
%\includegraphics[width=7cm]{chi_phi.eps}
%\caption{\label{chi1}  Behavior of the nonlinear term $-\chi$ in Eq. (\ref{pra8}) 
%as a function of $|\Phi|$
%for $\mu = 0$  and $\mu = 10$.}
%\end{center}
%\end{figure}

\begin{figure}[t]
\begin{center}
\includegraphics[width=8cm]{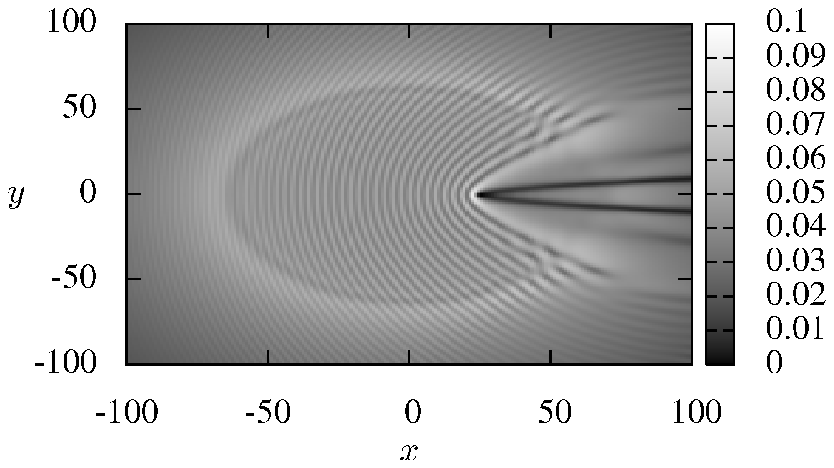}
\caption{\label{mu0} Numerical solution of Eq. (\ref{sb}) for the initial
profile given in Eq. (\ref{ic}) with $\sigma_1 = \sigma_2 = 8$. 
The density plot is shown for $t = 2$ and $\mu = 0$.
The dark ``V'' structure corresponds to a pair of oblique dark solitons and the oscillation
in front of the obstacle corresponds to the ``ship'' waves. 
An impenetrable disc of radius $r = 1$ is placed at $(25,0)$ in a BEC radially expanding from
the centre $(0,0)$.}
\end{center}
\end{figure}
\begin{figure}[t]
\begin{center}
\includegraphics[width=7cm]{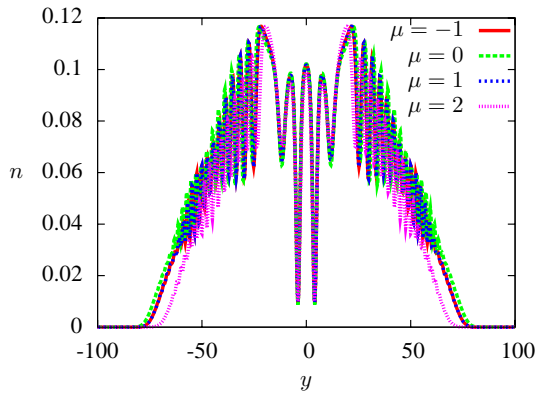}
\caption{\label{cuts} Cross sections of the density distributions for $t = 2$
at $x = 60$ for chemical potential $\mu = -1$, $\mu = 0$, $\mu = 1$ and $\mu = 2$.
The profile for different small $\mu$ is almost the same.}
\end{center}
\end{figure}

%%%%%%%%%%%%%%%%%%%%%%%%%%%%%%%%%%%%%%%%%%%%%%%%%%%%%%%%%%%%%%%%%%%%%%%%%%%%%%%%%%%%%%%%%%%%%%%%%%%%
\subsection{Discussion of the nonlinear term}

The form of the nonlinear term $\chi$ is crucial for the physics of the strongly interacting
Bose gas and makes all the difference in comparison to the GP approach. Despite the fact that
it is defined in the integral representation of Eqs. (\ref{pra8}), (\ref{pra4}), and (\ref{pra9}),
the interpretation of its asymptotic behavior is relatively simple.
First of all, this term can be expanded for low condensate density (i.e. for small $\Phi$) to get
\[
\chi\sim \chi(0) +\chi'(0)|\Phi|^2
\]
with $\chi'(0)<0$ \cite{jphys40_629}. The increasing behavior of this term with $|\Phi|^2$
reflects the repulsive nature of the interaction between the bosons.
The truncation after the quadratic term $|\Phi|^2$ leaves us with the Gross-Pitaevskii
Hamiltonian. In the opposite limit, i.e. for $|\Phi|\sim\infty$, we get
\[
\chi\sim const. |\Phi|^{-1}
.
\]
The behavior of $-\chi$ as a function of $|\Phi|^2$ is shown in Fig. \ref{chi1} for two different
values of the chemical potential $\mu$. The nonlinearity is obviously much weaker than that of the
GP equation. Moreover, it becomes weaker with increasing $\mu$ (i.e. increasing density of the 
Bose gas). This reflects the depletion of the condensate for increasing densities.
%\begin{figure}[t]
%\begin{center}
%\includegraphics[width=8cm]{chi_phi.eps}
%\caption{\label{chi1}  Behavior of the nonlinear term $-\chi$ in Eq. (\ref{pra8}) 
%as a function of $|\Phi|$
%for $\mu = 0$  and $\mu = 10$.}
%\end{center}
%\end{figure}

%%%%%%%%%%%%%%%%%%%%%%%%%%%%%%%%%%%%%%%%%%%%%%%%%%%%%%%%%%%%%%%%%%%%%%%%%%%%%%%%%%%%%%%%%%%%%%%%%%%%
\subsection{Dynamics of the macroscopic wave function}

The dynamics of the quantum state of the condensate is given by the nonlinear Schr\"odinger
equation:
%\[
%i\hbar \partial_t\Phi=H_\Phi\Phi
%\]
%describes the evolution of the macroscopic quantum state of the condensate.
%The dynamics of a Bose gas is described in the SB approach by the SB equation
%\begin{tinny}
\begin{eqnarray}\label{sb}
&{\displaystyle i \hbar \, \frac{\partial \Phi({\bf{r}},t)}{\partial t}~=~H_\Phi\Phi({\bf{r}},t)}&
\nonumber\\
&= {\displaystyle-\frac{J a^2}{6} \, \Delta \Phi({\bf{r}},t) \, + \,
J\left(
\alpha_1 - \alpha_2 \chi \right) \Phi({\bf{r}},t)} \, ,& 
\end{eqnarray}
%\end{tinny}
which will be called SIG equation.
This differential equation describes the expansion of the condensed part of a
bosonic cloud for a given initial state $\Phi({\bf{r}},0)$ at time $t=0$.
The spatial scale is given by scattering length $a_s \sim a$.  A
typical experimental value for $^{85}\text{Rb}$ atoms near a Feshbach resonance 
is $a_s \sim 200 \, \text{nm}$ \cite{jphys40_629}.

%\begin{figure}[h!!]
%\begin{center}
%\includegraphics[width=10cm]{SBvort.eps}
%\caption{\label{vortices}  Numerical solution of Eq. (\ref{sb}) for the initial
%profile given in Eq. (\ref{ic}) with $\sigma_1 = 16$ and $\sigma_2 = \sqrt{3} \sigma_1$. 
%The density plot is shown for $t = 2$ and $\mu = 0$.
%We observe the generation of pairs of vortices and antivortices past the obstacle.
%An impenetrable disc of radius $r = 1$ is placed at $(25,0)$ in a BEC radially expanding from
%the centre $(0,0)$.}
%\end{center}
%\end{figure}

%\begin{figure}[ht]
%\begin{center}
%\includegraphics[width=9cm]{mu0c.eps}
%\caption{\label{mu0} Numerical solution of Eq. (\ref{sb}) for the initial
%profile given in Eq. (\ref{ic}) with $\sigma_1 = \sigma_2 = 8$. 
%The density plot is shown for $t = 2$ and $\mu = 0$.
%The dark ``V'' structure corresponds to a pair of oblique dark solitons and the oscillation
%in front of the obstacle corresponds to the ``ship'' waves. 
%An impenetrable disc of radius $r = 1$ is placed at $(25,0)$ in a BEC radially expanding from
%the centre $(0,0)$.}
%\end{center}
%\end{figure}
%\begin{figure}[ht]
%\begin{center}
%\includegraphics[width=8cm]{sol.eps}
%\caption{\label{cuts} Cross sections of the density distributions for $t = 2$
%at $x = 60$ for chemical potential $\mu = -1$, $\mu = 0$, $\mu = 1$ and $\mu = 2$.
%The profile for different small $\mu$ is almost the same.}
%\end{center}
%\end{figure}

%%%%%%%%%%%%%%%%%%%%%%%%%%%%%%%%%%%%%%%%%%%%%%%%%%%%%%%%%%%%%%%%%%%%%%%%%%%%%%%%%%%%%%%%%%%%%%%%%%%%
\section{Numerical solution and results}

To study the evolution of an initial state $\Phi({\bf r},t=0)$ with Eq. (\ref{sb}) we assume a
two-dimensional (2D) situation, where the expansion is possible in the $x,y$ direction but not
in the $z$ direction. This is a typical case in which the trapping potential, after the formation
of the BEC, is switched off in two directions but a strongly confining potential is kept in the 
perpendicular direction. Then we choose for the initial state a 2D Gaussian function by
\begin{equation}\label{ic}
\Phi(x,y,t=0)~=~\sqrt{\sigma_1} \, \exp \left(-0.5 \, \frac{(x^2 + y^2)}{\sigma_2^2}\right)
\ ,
\end{equation}
where $\sigma_1$ and $\sigma_2$ are parameters which determine the shape of the Gaussian.
%In general, the macroscopic state $\Phi({\bf{r}},t)$ is zero at the boundary.

\begin{figure}[t]
\begin{center}
\includegraphics[width=8cm]{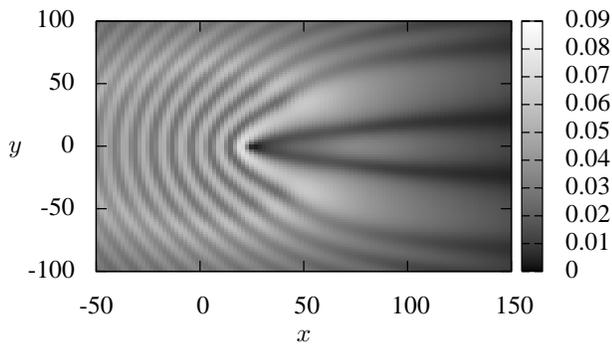}
\caption{\label{mu10_7} Numerical solution of Eq. (\ref{sb}) for the initial
profile given in Eq. (\ref{ic}) with $\sigma_1 = \sigma_2 = 8$. 
The density plot is shown for $t = 6$ and $\mu = 10$.
The dark ``V'' structure corresponds to a pair of oblique dark solitons and the oscillation
in front of the obstacle corresponds to the ``ship'' waves. 
An impenetrable disc of radius $r = 1$ is placed at $(25,0)$ in a BEC radially expanding from
the centre $(0,0)$.}
\end{center}
\end{figure}
\begin{figure}[t]
\begin{center}
\includegraphics[width=8cm]{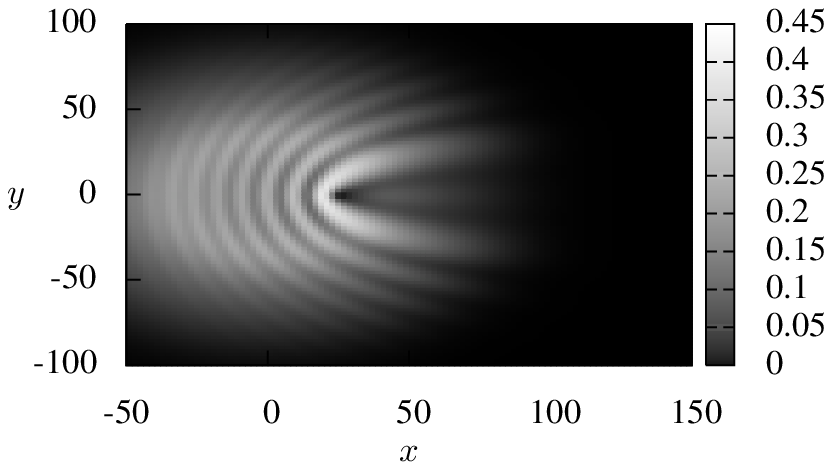}
\caption{\label{mu20_7} Numerical solution of Eq. (\ref{sb}) for the initial
profile given in Eq. (\ref{ic}) with $\sigma_1 = \sigma_2 = 8$. 
The density plot is shown for $t = 6$ and $\mu = 20$.
The dark ``V'' structure corresponds to a pair of oblique dark solitons and the oscillation
in front of the obstacle corresponds to the ``ship'' waves. 
An impenetrable disc of radius $r = 1$ is placed at $(25,0)$ in a BEC radially expanding from
the centre $(0,0)$. 
We see that for bigger $\mu$, the expansion of the BEC is much slower than in the
case illustrated in Fig. \ref{mu10_7}.}
\end{center}
\end{figure}

We apply a 2D finite-difference method (Crank-Nicolson method) combined with a 
split-step method to solve equation (\ref{sb}) numerically.
Schematically this reads
\begin{eqnarray}
  &\Phi(x,y,t)=e^{-i H t} \Phi(x,y,0)&\nonumber\\
  &~\approx~
  e^{-i V \frac{\delta t}{2}} e^{-i T \delta t} 
  e^{-i V \frac{\delta t}{2}} \Phi(x,y,0)\approx&\\
  &e^{-i V \frac{\delta t}{2}} 
  \mathcal{CN}\left[ y,\frac{\delta t}{2} \right] 
  \mathcal{CN}\left[ x,\delta t \right]
  \mathcal{CN}\left[ y,\frac{\delta t}{2}  \right]
                e^{-i V \frac{\delta t}{2}} 
                \Phi(x,y,0)&\nonumber
\end{eqnarray}
where $V$ and $T$ are the potential and kinetic term, respectively,
$\delta t$ is the time step and $\mathcal{CN} [x,\delta t]$ is the
action of the Crank-Nicolson method for the kinetic term on the $x$ direction. 
This approach is accurate up to $\mathcal{O}(\delta t^3)$,
and good results are obtained for sufficiently small time step $\delta t$.
We use a Gauss-Hermite quadrature to calculate at each time step the integrals 
$\chi$ and $Z({\bf{r}})$ in (\ref{pra8}) and (\ref{pra4}), respectively,
and noticed that the Gauss-Hermite integration converges already for only 
nine points.

In our simulations, we use as boundary condition $\Phi({\bf{r}}) = 0$ at
the boundary of a square region.
An impenetrable obstacle is implemented by assuming that the order parameter
vanishes %$\Phi({\bf{r}}) = 0$ 
inside a circle of radius $r=1$.
The obstacle can also be implemented as %$V({\bf{r}})$ 
a strong repulsive Gaussian potential
$V({\bf{r}}) = U_0 \exp (-(x^2 + y^2)/2 \kappa)$
with the chemical potential given by $\mu({\bf{r}}) = \mu - V({\bf{r}})$.
Both conditions give similar results.

It was shown in \cite{frisch} that the flow of a dilute BEC 
past an obstacle generates vortices and antivortices at subsonic velocity.
Increasing the flow velocity, the frequency of vortices generation also increases.
For large enough flow velocity, the vortices generation is so often,
that the distance between them becomes less than their radial size and 
it takes a long time for their separation from each other
at finite distance from the obstacle \cite{winiecki_prl,winiecki,prl97_180405,anatoly}.
We study this transition from generation of vortices to generation of dark solitons 
in a dense Bose gas, considering different profiles
of the initial state (different values of $\sigma_1$ and $\sigma_2$ in Eq. (\ref{ic})),
which corresponds to faster expansion for a small initial profile and 
to slower expansion, for bigger initial profile.

We observe the emergence of pairs of vortices and antivortices 
using $\sigma_1 = 16$ and $\sigma_2 = \sqrt{3} \sigma_1$ in Eq. (\ref{ic}) 
(cf. Fig. \ref{vortices}). 
This is similar to what was observed in recent experiments \cite{neely10}.

We study the case of faster expansion of the dense BEC for the initial state 
in Eq. (\ref{ic}) with $\sigma_1 = \sigma_2 = 8$
and different values of $\mu$, 
namely $\mu = -1$, $\mu = 0$, $\mu = 1$, $\mu = 2$, $\mu = 10$ and $\mu = 20$.
We observe in Fig. (\ref{mu0} - \ref{mu20_7}) a pair of oblique dark solitons 
past the obstacle and linear waves, as in the case of GP \cite{prl97_180405}. 
For small values of chemical potential, $\mu = -1$, $\mu = 0$, $\mu = 1$, $\mu = 2$, 
the pattern (position, slope and amplitude of the soliton) is almost the same
(cf. Fig. \ref{cuts}).
However, we find that the velocity of the expansion of the cloud becomes smaller 
for an increasing chemical potential (see Fig. \ref{mu10_7} and Fig. \ref{mu20_7}). 

\begin{figure}[t]
\begin{center}
\includegraphics[width=7cm]{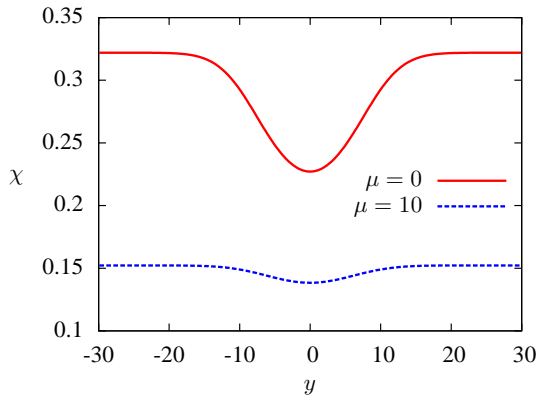}
\caption{\label{chi} Cross section at $x=0$ of the nonlinear term $\chi$ in Eq. (\ref{pra8}) for 
$\mu = 0$ and $\mu = 10$. It goes to zero as we increase $\mu$.}
\end{center}
\end{figure}

\begin{figure}[t]
\begin{center}
\includegraphics[width=7cm]{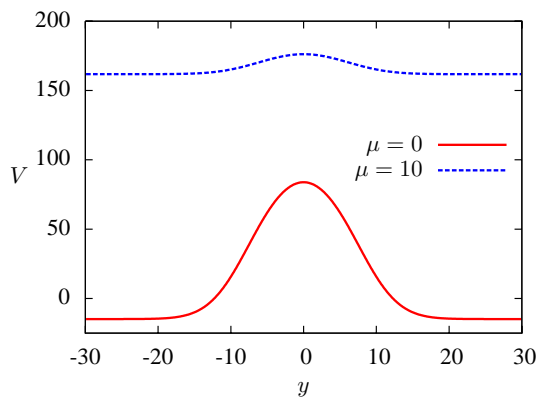}
\caption{\label{Vfull} Cross section at $x=0$ of the potential term in Eq. (\ref{sb}) for $\mu = 0$ 
and $\mu = 10$. It becomes flat as we increase $\mu$. It makes the expansion of the cloud slower
for big $\mu$.}
\end{center}
\end{figure}

We also study the effect of the nonlinear term $\chi$ in Eq. (\ref{pra8}) 
with the full potential of Eq. (\ref{sb}) for different values of $\mu$ on the dynamics.
Since $\chi$ becomes smaller as we increase $\mu$ (see Fig. \ref{chi}), the expansion 
velocity of the cloud is also reduced with increasing $\mu$.
Correspondly, the effective potential term of the Hamiltonian is strongly repulsive for small $\mu$ 
and becomes flat for bigger $\mu$ (see Fig. \ref{Vfull}). 
%\begin{figure}[h!!]
%\begin{center}
%\includegraphics[width=8cm]{chi.eps}
%\caption{\label{chi} Cross section at $x=0$ of the nonlinear term $\chi$ in Eq. (\ref{pra8}) for 
%$\mu = 0$ and $\mu = 10$. It goes to zero as we increase $\mu$.}
%\end{center}
%\end{figure}
%
%\begin{figure}[h!!]
%\begin{center}
%\includegraphics[width=8cm]{Vfull.eps}
%\caption{\label{Vfull} Cross section at $x=0$ of the potential term in Eq. (\ref{sb}) for $\mu = 0$ 
%and $\mu = 10$. It becomes flat as we increase $\mu$. It makes the expansion of the cloud slower
%for big $\mu$.}
%\end{center}
%\end{figure}

Although the nonlinearity of the SIG equation is more complex than its counterpart in
the GP equation, the flow past an obstacle shows qualitatively the same patterns as for the GP equation:
it is characterized by oblique dark solitons and by linear waves. Our study suggests that these effects 
shall well appear in a strongly interacting gas too and may be observed experimentally.  

\section{Conclusion}
We have studied the two-dimensional expansion of a strongly interacting BEC 
in the presence of an obstacle.
The strong interaction was treated within the slave-boson approach, which leads to a nonlinear
Schr\"odinger equation with a special nonlinearity, different from the Gross-Pitaevskii term.
We solved numerically this equation % for an expanding condensate that passes an obstacle
and analyzed the contribution of the nonlinearity.
Similar to what was found in previous studies of the GP equation \cite{frisch,prl97_180405}, 
we observed the 
pairwise generation of vortices and antivortices (at low velocity) and
of shock waves (at high velocity), respectively.
We noticed that bigger values of the chemical potential (i.e. for higher densities of the Bose gas) 
the expansion is slower due to a reduced nonlinear term. The characteristic features of the 
SIG equation in terms of a generation of vortices, oblique dark solitons and ship waves are similar to
those described by the GP.

\begin{acknowledgments}
This work was supported by Coordena\c c\~ao de Aperfei\c coamento de Pessoal 
de N\'ivel Superior (CAPES) and Deutscher Akademischer Austausch Dienst (DAAD).
AG also thanks Brazilian funding agencies FAPESP and CNPq.
\end{acknowledgments}

%%%%%%%%%%%%%%%%%%%%%%%%%%%%%%%%%%%%%%%%%%%%%%%%%%%%%%%%%%%%%%%%%%%%%%%%%%%%%%%%%%%%%%%%%%%%%
%\newpage

\end{document}